\documentclass{article}
\def\dslash {\partial\!\!\!/}
\newcommand{\be}{\begin{equation}}
\newcommand{\ee}{\end{equation}}
\newcommand{\bear}{\begin{eqnarray}}
\newcommand{\ear}{\end{eqnarray}}
\newcommand{\ba}{\begin{eqnarray*}}
\newcommand{\ea}{\end{eqnarray*}}
\renewcommand{\theequation}{\arabic{section}.\arabic{equation}}

\newcommand{\no}{\noindent}

\newcommand{\dt}{\mbox{\boldmath$:$}}

\begin{document}
\title{Quantum Field Lagrangian model for charge density waves in one-dimensional systems at finite temperature : A Thermofield Dynamics Approach}
\author{L. V. Belvedere$^1$, R. L. P. G. Amaral$^{1,2}$ and A. F. Rodrigues$^3$\\
\small{$1$ Instituto de F\'{\i}sica}\\
\small{Universidade Federal Fluminense}\\
\small{Av. Litor\^anea S/N, Boa Viagem, Niter\'oi, CEP. 24210-340}\\
\small{Rio de Janeiro - Brasil}\\
\small{$2$ Center for Theoretical Physics, Massachusetts Institute of Technology, Cambrige, MA 02138, USA}\\
\small{$3$ ICEx - Instituto de Ci\^encias Exatas}\\
\small{Universidade Federal Fluminense /PUVR- Volta Redonda, CEP. 27213-415}\\
\small{RJ-Brasil}\\
}
\date{\today}
\maketitle
\begin{abstract}
We consider the Thermofield Dynamics bosonization to perform a field theory analysis of the effective Lagrangian model for incommensurate charge density waves (ICDW) in one-dimensional systems at finite temperature. The phonon degree of freedom is  carryied by a dynamical phase field, contributing to the quantum dynamics and symmetry related features of the ICDW phenomenon. The electron chiral density and the phase of the phonon field condensate as a thermal soliton, carrying the symmetry under the linked electron-phonon $U_e^5(1) \otimes U_{ph}(1)$ global transformations. Using the Gell'Mann-Low formula for finite temperature, the perturbative series of the phonon thermal correlation function is obtained. Due to the  electron-phonon charge selection rule we obtain for the  thermal vacuum expectation value for the order parameter $\langle 0 (\beta ) \vert \Phi \vert 0 (\beta ) \rangle = 0$, in accordance with the cluster decomposition property of the corresponding correlation function. This reflects the fact that the quantum description of the ICDW corresponds to a local charge transport through the lattice which is accomplished by an electron-lattice energy redistribution, which accounts for a thermal dynamical mass gap generation. The electron-phonon coupling can be rewritten in terms of a mass operator for the ''physical'' fermion operator $\Psi$ such that $\langle 0 (\beta )\vert \bar\Psi (x)\Psi (x) \vert 0 (\beta) \rangle \neq 0$, without the breakdown of the linked electron-phonon symmetry. 
\end{abstract}

\section{Introduction}
\setcounter{equation}{0}

An impressive effort has been made by many physicists along several years to understand the underlying properties of quantum field theories in two-dimensions \cite{AAR}, as well as to picture these models as theoretical laboratories to obtain insight into more realistic four-dimensional field theory. More recently, attention have been paid to apply these models to low-dimensional condensed matter systems(\cite{Fradkin}-\cite{BAQ3}).

The formation and acceleration of ICDW in low-dimensional systems \cite{Gruner,SuSakita,Saka2,Krive,Ishikawa} has been described in terms of an underlying chiral symmetry of the electron-phonon system. Since the CDW transport  features can be pictured as essentially a one-dimensional phenomenon, even for two- or three-dimensional structures \cite{Gruner}, quantum field theory methods in $1 + 1$ dimensions (specially bozonization of fermions) have been applied to study the effective Lagrangian models for the charge density waves 
systems (\cite{SakitaShizuya}-\cite{BAQ3}).

The interest for CDW transport phenomena within field theoretical settings in condensed matter physics has gained recently new endorsements, and remains attracting attention. In the study of carbon nanotubes \cite{Sasaki}, two-dimensional field theory, similar to that one dealt with here, raises the question, whether a phonon field  needs to be taken into account for the proper computation of finite temperature contributions to the electrical behavior and vacuum structure of carbon nanotubes. Also, mixed states in ICDW with three dimensional order have been described in \cite{Hayashi} via a formalism that parallels the chiral decomposition employed in \cite{SakitaShizuya}.

The field theoretical models have been used \cite{Ya} with Lorentz-like variables for the description of the quantum Hall effect in one-dimensional systems. The holographic CDW formalism has been developed recently \cite{Aperis} which leads to a commensurable CDW on the frontiers and which is described by a model quite similar to that one discussed here. In a different perspective, massless modes in  cosmic strings are treated by field theoretical methods keeping close analogy to condensed matter systems descriptions \cite{Hind}. This  approach leads to two-dimensional  fermions interacting with complex scalar fields. 


The bosonization of fermions has proven in the past to be a very useful technique for solving quantum field theoretic models in $1 + 1$ dimensions \cite{AAR}. In a previous paper \cite{BAQ1} a field theory analysis for the effective Lagrangian approach to ICDW in quasi-one-dimensional systems was presented using the standard bosonization formalism. 

The effective electron-phonon Lagrangian for the ICDW exhibits symmetry under the linked electron-phonon $U^5_e(1) \otimes U_{ph}(1)$ transformations. This extended symmetry transformation that couples the electron and phonon fields acts as the underlying mechanism that allows for the ICDW to emerge as a collective excitation of the electron-phonon system. The description of the dynamical collective phenomena of the electron-phonon system, and its related symmetry aspects, is performed using an approximation in which the fluctuations of the phonon amplitude mode are disregarded and only the fluctuations of the phase of the order parameter contributes to the quantum dynamics. The amplitude of the phonon field is frozen, which we call dynamical phonon phase approach. This provides a good approach for describing low energy processes. In this case, instead of considering both the amplitude and the phase of the order parameter as slowly varying fields \cite{Saka2}, we assume that!
  only the space-time variations of the amplitude can be neglected. This approximation affects deeply the description of the ICDW phenomenon and enables the ICDW to appear as a collective excitation of the combined motion of both the electrons and lattice ions. The phase of the phonon field and the phase of the bosonized electron chiral density condensate as a ``soliton'' order parameter, which is neutral under the linked $U^5_e(1) \otimes U_{ph}(1)$ global charge transformation. The electron-phonon condensate leads to a periodic  sine-Gordon potential that represents a dynamically generated mass gap without the corresponding chiral symmetry breakdown.

The purpose of the present paper is to discuss the dynamical gap generation and the role played by the electron-phonon linked symmetry at finite temperature. To this end we shall follow the same approach of refs. \cite{BAQ1,BAQ2,BAQ3}, to provide a brief discussion of the Lagrangian  description of ICDW phenomenon at finite temperature using the bosonization of Fermi fields within the thermofield dynamics formalism  \cite{TField}, the ``thermofield bosonization''  introduced in refs. \cite{ABR1,ABR2,ABR3}. Using the dynamical phase phonon approximation and within the finite temperature context, the ICDW can be pictured as an electron-phonon collective phenomenon in which the charge wave propagates together with a topological thermal soliton carrying a topologically conserved charge. The linked $U_{ph}(1) \otimes U^5_{e}(1)$ selection rule associated with this thermal collective excitation ensures the appearance of a  energy gap without the breakdown of the corresponding symm!
 etry.

The paper is organized as follows: In Section $2$ we present the formal aspects of the thermofield dynamics approach for the effective Lagrangian of the ICDW using the thermofield bosonization point of view. In Section $3$ we use the perturbative expansion to obtain the phonon two-point function. Some details of the computations are presented in  Appendix A. Also, we obtain the electron-phonon selection rule that ensures for the thermal vacuum expectation value for the phonon field  $\langle 0 (\beta ) \vert \Phi  \vert 0 (\beta ) \rangle = 0$, in accordance with the cluster decomposition for the thermal phonon two-point function. Finally, in section $4$ we present the conclusion and final remarks concerning the dynamical mass gap generation and the ''anomaly'' effect due to the introduction of an external electric field.

\section{Thermofield Dynamics Approach}
\setcounter{equation}{0}

The ''fenomenological'' effective Lagrangian model for ICDW  is obtained from the  Fr$\ddot{\mbox{o}}$hlich Lagrangian \cite{Froh} of interacting electron-phonon system by  linearizing the electron espectrum near the Fermi level, taking into account incommensurability by neglecting terms involving the wave-function phase factors, that depends on the Fermi momenta  $Q = 2 p_F$ (which is incommensurate with lattice spacing),  and neglecting the acoustic part of the phonon field \cite{Rice,Krive,SuSakita} is given by  \footnote{Our conventions are: $\hbar = c = 1$ 
$$
g^{\mu \nu} = \pmatrix{1 & 0 \cr 0 & 1 } , \epsilon^{\mu \nu} = \pmatrix{ 0 & 1 \cr - 1 & 0} ,\gamma^0 = \pmatrix{ 0 & 1 \cr 1 & 0 } ,\gamma^1 = \pmatrix{ 0 & 1 \cr -1 & 0 } , \gamma^5 = \gamma^0 \gamma^1 ,\gamma^\mu \gamma^5 = \epsilon^{\mu \nu} \gamma_\nu ,$$
$$
 \dslash = \gamma^\mu \partial_\mu , (\partial_\mu \phi)^2 \equiv \partial_\mu \partial^\mu\,,\, \psi = \pmatrix{\psi_\ell \cr \psi_r}\,,\,\bar\psi = \psi^\dagger \gamma^0\,.
$$
}

\be\label{L}
{\cal L}  = i\,\psi^\dagger_\ell \big ( \partial_t - v_F \partial_x \big ) \psi_\ell  + i\,\psi^\dagger_r  \big ( \partial_t + v_F \partial_x \big ) \psi_r (x) + \lambda_0 \Phi^\ast \big ( - \partial^2_t + v^2_Q \partial^2_x - \omega_0^2 \big ) \Phi + G_0 \big ( \Phi \psi^\dagger_r  \psi_\ell + \Phi^\ast \psi^\dagger_\ell \psi_r \big )\,,
\ee

\no where $\lambda_0$ is the linear density of the ion masses (which for simplicity we shall consider equal to one) and $G_0$ is the electron and lattice phonon modes scattering coupling constant. Here $\omega_0$ is the optical phonon  frequency $\omega (Q)$ at $Q = 2 p_F$ and $v^2_Q = \omega_0^2 / 2$.  Defining $x^\mu = (x^0 , x^1) \doteq (t , x / v_F )$, $\psi^\prime = v_F^{1 / 2} , \Phi^\prime = v_F^{1 / 2} \Phi$, rescaling the electron-phonon coupling constant $G = v_F^{ - 1 /2} G_0$, and streamlining the notation by dropping ``primes'' everywhere, the electron-phonon Lagrangian (\ref{L}) can be written as a two-dimensional ``relativistic'' field model

\be
\label{L1}
{\cal L} = \overline{\psi} i \gamma^\mu \partial_\mu \psi - \Phi^\ast \partial^2_a \Phi - \omega_0^2 \Phi^\ast \Phi + G \big ( \Phi \psi^\dagger_r \psi_\ell + \Phi^\ast \psi^\dagger_\ell \psi_r \big )\,,
\ee

\no where
$$
\partial^2_a = \partial_0^2 - a^2 \partial_1^2
$$
\no with $a = v_Q / v_F$. In what follows we shall discuss the ``Lorentz invariant'' case in which $a = 1$ and $\partial^2_a$ becomes the usual D'Alembert operator $\partial^2 = \partial_\mu \partial^\mu$. The Lagrangian (\ref{L1}) is invariant under the linked electron-phonon $U^5_e (1) \otimes U_{ph} (1)$ global transformations
$$
\psi^\prime (x) = e^{\textstyle i \gamma^5 \frac{\kappa}{2}} \psi (x)\,,
$$
$$
\Phi^\prime (x) = e^{\textstyle -\,i \kappa} \Phi (x)\,.
$$

The effective Lagrangian (\ref{L}) is given in terms of the phonon field $\Phi (x)$ and the Fermi field $\psi (x)$, from which the Hilbert space ${\cal H}$ is builded . The starting point of the thermofield dynamics formalism is the doubling of the Hilbert space \cite{TField,TField1,TField2,TField3,Ojima} by introducing the ``tilde'' fields $\widetilde\Phi (x)$ and $\widetilde\psi (x)$. At $T = 0$,  the tilde fields lives in the Hilbert space $\widetilde{\cal H}$ and are independent identical copies of the original fields $\Phi (x) , \psi (x) \in {\cal H}$. The total Lagrangian is given by

\be\label{Lhat}
\widehat{\cal L} = {\cal L} - \widetilde{\cal L}\,,
\ee

\no where the ``tilde'' operation in (\ref{Lhat}) is defined by $\widetilde{(c\Phi(x))} = c^\ast \widetilde\Phi (x)$. The Lagrangian (\ref{Lhat}) is invariant under the extended  $\big [ U^5_e (1) \otimes U_{ph} (1) \big ] \times \big [ \widetilde{U}^5_e (1) \otimes \widetilde{U}_{ph} (1) \big ]$ transformations.

We shall assume that only the space-temporal variations of the phonon amplitude $\rho (x)$ are neglected, $\partial \rho \approx 0$ \cite{BAQ1,BAQ2}, such that the description of dynamical collective phenomenon is in terms of the dynamics of the order-parameter phase only. This approach is expected to be consistent in describing finite temperature effects as long as the temperature is not so high in order to excite the amplitude modes of the phonon field. In the range of temperatures ($\beta^{- 1} = K T$)  smaller than the range of the phonon optical frequencies,  the  amplitude fluctuations do not play an importante role and the collective phenomenon can be described in terms of the dynamics of the phase only \cite{Gruner}. In this case the phase $\varphi (x)$  is considered a dynamical degree of freedom, that contributes to the quantum dynamics of the ICDW system. This approximation displays several aspects of the linked electron-phonon  symmetry that enables the ICDW to a!
 ppear as a collective transport phenomenon due to the combined motion of both electrons and lattice ions.

To begin with, at $T = 0$, we shall introduce the phonon field in terms of the infrared regularized Wick normal-ordered exponential of the  field phase $\varphi (x)$ (for simplicity in what follows we shall consider the amplitude $\rho (x) = \rho_0$ and $\alpha$ defines the value of the scale dimension  of the exponential operator) \cite{BAQ1,BAQ2}

\be\label{phi}
\Phi (x) =  \rho_0 \big (\, \mu \,\big )^{\textstyle \frac{\alpha^2}{4 \pi}} \dt  e^{\textstyle\,i \alpha  \varphi (x)} \dt \,,
\ee
\be\label{phitilde}
\widetilde\Phi (x) =  \rho_0 \big (\, \mu \,\big )^{\textstyle \frac{\alpha^2}{4 \pi}} \dt  e^{\textstyle\,i \alpha  \widetilde\varphi (x)} \dt \,.
\ee

\no The bosonized Fermi field is given in terms of the  Mandelstam \cite{M} formula
\be\label{psi}
\psi (x) = \big ( \mu \big )^{ \frac{\gamma^2}{8 \pi}}\,\dt e^{\textstyle\,i\,\gamma^5\,\frac{\gamma}{2} \eta (x)\,+\,\frac{2 \pi}{\gamma}\,\int_x^\infty\,dz^1\,\partial_0 \eta (z)}\,\dt\,,
\ee
\be\label{psitilde}
\widetilde\psi (x) = \big ( \mu \big )^{ \frac{\gamma^2}{8 \pi}}\,\dt e^{\textstyle\,i\,\gamma^5\,\frac{\gamma}{2} \widetilde\eta (x)\,+\,\frac{2 \pi}{\gamma}\,\int_x^\infty\,dz^1\,\partial_0 \widetilde\eta (z)}\,\dt\,,
\ee

\no and the bosonized fermion chiral densities \cite{ABR1,ABR2,ABR3,Swieca,M} are  given by 

\be
\vdots\,\psi_\ell^\ast (x) \psi_r (x) \vdots = J (x) = \big ( \frac{\mu}{2 \pi} \big )^{\frac{\gamma^2}{4 \pi}} \dt e^{\,i\,\gamma\,\eta (x)} \dt\,,
\ee
 
\be
\vdots\,\widetilde\psi_\ell^\ast (x) \widetilde\psi_r (x) \vdots = \widetilde{J} (x) = \big ( \frac{\mu}{2 \pi} \big )^{\frac{\gamma^2}{4 \pi}} \dt e^{\,i\,\gamma\,\widetilde\eta (x)} \dt\,.
\ee

\no The canonical case corresponds to $\gamma = 2 \sqrt \pi$. The Lorentz spin is $S_\psi = (\gamma / 2)(2 \pi / \gamma ) = 1 / 2$.  Notice that the bosonized expression (\ref{psitilde}) for the Fermi field $\widetilde\psi$, as well as for the phonon fields (\ref{phit}), are not obtained from the corresponding expression (\ref{psi}) for the field $\psi$ and expression (\ref{phi}) for the field $\Phi$,  by the ``tilde conjugation operation'' of the Wick-ordered exponential operators. We shall return to this point later. The dots $\dt (\,\cdots\,) \dt$ mean normal ordering defined with respect to the positive- and negative-frequency parts $\varphi^{(\pm)} (x)$ and the parameter $\mu$ is an infrared (IR) regulator  \footnote{The regularization is needed in order to have a well defined exponential field operator. This is due to the fact that the two-dimensional free massless scalar field is not well defined by itself due to the infrared divergence of the corresponding two-point !
 function,
$$
\langle 0 \vert \varphi (x) \varphi (0) \vert 0 \rangle = - \frac{1}{4 \pi}\,\ln \{\,-\,\mu^2\,x^2\,-\,i \epsilon \}\,.
$$
\no At $T = 0$ one has an infrared cut-off independent two-point function,
$$
\langle 0 \vert \Phi (x) \Phi^\ast (0) \vert 0 \rangle = \big ( - x^2 - i \epsilon \big )^{ -\,\frac{\alpha^2}{4 \pi}}\,.
$$}. Computing the kinetic and mass terms of the phonon field in the Lagrangian (\ref{L1}) using the Wilson short-distance expansion \cite{BAQ1} (here VEV stands for the vacuum expectation value of the corresponding point-splitted operator product and ${Z}$ is a ''wave function'' renormalization constant),

$$
\dt \partial^\mu \Phi (x) \partial_\mu \Phi^\ast (x) \dt \doteq \lim_{\varepsilon \rightarrow 0}\,\frac{1}{2} \Big \{ Z^{- 1} (\varepsilon ) \big [ \partial_\mu \Phi (x + \varepsilon ) \partial^\mu \Phi^\ast (x) + \partial_\mu \Phi (x) \partial^\mu \Phi^\ast (x + \varepsilon ) \big ] - V. E. V \Big \}
$$

\no and similarly for the phonon ``mass term'', we get for the corresponding total free Lagrangian piece \cite{BAQ1}

\be
\widehat{\cal L}^{(0)}_{ph} (x) = \frac{\rho}{2} \dt (\partial_\mu \varphi (x) )^2 \dt\,-\,\frac{\rho}{2} \dt (\partial_\mu \widetilde\varphi (x) )^2 \dt\,,
\ee

\no where $\rho = \alpha\rho_0\omega_0$ \cite{BAQ1}. Taking this into account, performing a new rescaling $\psi^\prime = \rho^{- 1} \psi$ of the electron field, using the bosonized correspondence for the fermionic kinetic term

$$
i \bar \psi (x) \gamma^\mu \partial_\mu \psi (x) = \frac{1}{2} \dt (\partial_\mu \eta (x) )^2 \dt\,,
$$

\no and streamlining the notation by dropping ``primes'' everywhere, we obtain  the bosonized effective Lagrangians

\be\label{LBos}
{\cal L} = \frac{1}{2} \dt (\partial_\mu \eta (x) )^2 \dt + \frac{1}{2} \dt (\partial_\mu \varphi (x) )^2 \dt + G \Big ( \frac{\mu}{2 \pi} \Big )^{\textstyle \frac{\gamma^2}{4 \pi}} \, \dt \cos [ \gamma \eta (x) - \alpha \varphi (x) ] \dt \,,
\ee
\be\label{tLBos}
\widetilde{\cal L} = \, \frac{1}{2} \dt (\partial_\mu \widetilde\eta (x) )^2 \dt + \frac{1}{2} \dt (\partial_\mu \widetilde\varphi (x) )^2 \dt
+ G \Big ( \frac{\mu}{2 \pi} \Big )^{\textstyle \frac{\gamma^2}{4 \pi}} \,\dt \cos [ \gamma \widetilde\eta (x) - \alpha \widetilde\varphi (x) ] \dt \,.
\ee

\no In the bosonized version of the model, the electron-phonon transformations are mapped into the field shifts 
$$
 \eta (x) \rightarrow  \eta (x) + \frac{\kappa}{\gamma} \,\,,\,\, \varphi (x) \rightarrow \varphi (x) +  \frac{\kappa}{\alpha}\,.
$$
\no  The currents that generate the $U^5_e(1)$ and $U_{ph}(1)$ transformation are obtained from the Wilson expansions
$$
j^{\mu 5}_e (x) \doteq \vdots \bar\psi (x) \gamma^\mu \gamma^5 \psi (x) \vdots = \frac{1}{2} {\lim_{{\varepsilon \rightarrow 0}\atop{\varepsilon^2 < 0}}} \,f (\varepsilon ) \big \{ \bar\psi (x + \varepsilon) \gamma^\mu \gamma^5 \psi (x)\,-V. E. V. \big \}\,,
$$
$$
j^\mu_{ph} (x) = \frac{i}{2} {\lim_{{\varepsilon \rightarrow 0}\atop{\varepsilon^2 < 0}}} F (\varepsilon ) \big \{ \Phi^\ast (x + \varepsilon ) \partial^\mu \Phi (x) - \partial^\mu \Phi^\ast (x) \Phi (x + \varepsilon ) \,- V.E.V. \big \},
$$

\no  similarly for the tilde currents, and we get 

\be\label{jel}
j^{\mu 5}_e (x) = \epsilon^{\mu \nu } j_{e \nu} (x) = - \frac{\gamma}{2 \pi} \partial^\mu \eta (x)\,,
\ee

\be\label{jeltilde}
\widetilde j^{\mu 5}_e (x) = + \frac{\gamma}{2 \pi} \partial^\mu \widetilde\eta (x)\,,
\ee

\be\label{jph}
j^\mu_{ph} (x) = - \frac{\alpha}{2\pi} \partial^\mu \varphi (x)\,,
\ee

\be\label{jphtilde}
\widetilde j^\mu_{ph} (x) = + \frac{\alpha}{2\pi} \partial^\mu \widetilde\varphi (x)\,.
\ee

\no Although the currents $J^{\mu 5}_e$ and $j^\mu_{ph}$ are not conserved  we can define the current
\be
{\cal I}^\mu (x) \doteq \frac{\alpha}{\gamma} j^{\mu 5}_e (x) + \frac{\gamma}{\alpha} j^\mu_{ph} (x)\,,
\ee
\no such that 
\be
\partial_\mu {\cal I}^\mu (x) =  0\,,
\ee
\no as required by the equations of motion of the Bose fields $\eta$ and $\varphi$. Similary one has for the tilde current $\partial^\mu \widetilde{\cal I}^\mu = 0$.

It should be stressed that the equivalence between the mass operator $\bar\psi (x)\psi (x)$ and the sine-Gordon operator $:cos \gamma \eta (x):$ is obtained on the basis of the short distance expansion valid for massless free theory, that is, in order to the ``massive'' theory has the free model as 
short distance fixed point, one must requires for  the scale dimension of the mass operator \cite{Swieca}  
$$ 
D_{_{\bar\psi\psi}} = \gamma^2 / 4 \pi < 2\,.
$$
\no If this inequality is violated, for short distances one will be driven away from the fixed point \cite{Swieca}. In order to discard the Thirring coupling in (\ref{psi}) \cite{Swieca,Col,M} we shall consider the canonical case
$$
\gamma^2 = \alpha^2 = 4 \pi\,.
$$

Let us comment the physical interpretation for the fact that the bosonized expression (\ref{psitilde}) for the field $\widetilde\psi$ (as well as for $\widetilde\Phi$) is not obtained from the corresponding expression (\ref{psi})  for the field $\psi$ by the ``tilde conjugation operation''. In  Thermofiled Dynamics formalism, the fictitious ``tilde''  system should be an identical copy of the system under consideration, which implies that the field $\widetilde\psi$ should be an identical copy of $\psi$, and thus carrying the same charge and chirality quantum numbers. At $T = 0$, we have the expression for the two-dimensional free massless scalar field \cite{ABR1,ABR2}
$$
\varphi (x) = \int_{- \infty}^{+ \infty} (dp) \big ( f_p (x) a (p^1) + f^\ast_p (x) a^\dagger (p^1) \big )\,,
$$
$$
\widetilde\varphi (x) = \int_{- \infty}^{+ \infty} (dp) \big ( f^\ast_p (x) \tilde a (p^1) + f_p (x) \tilde a^\dagger (p^1) \big )\,,
$$

\no where

$$
(dp) = \frac{d p^1}{\sqrt{4 \pi \vert p^1 \vert}}\,,
$$
\no and 
$$
f_p (x) = e^{ \textstyle - i p^\mu x_\mu}\,.
$$
\no The two-dimensional free and massless scalar field can be decomposed in terms of the left- and right-moving fields ($x^\pm = x^0 \pm x^1$ ),
$$
\phi (x) = \phi (x^+) + \phi (x^-)\,,
$$
\no and the corresponding two-point functions are computed with respect to the Fock vacuum state $\vert 0 , \widetilde 0 \rangle = \vert 0 \rangle \otimes \vert \widetilde 0 \rangle$, 

$$
\langle 0  \vert \varphi (x^\pm) \varphi (0) \vert  0 \rangle = - \frac{1}{4 \pi} \ln [ i \mu ( x^\pm - i x^0 \epsilon)]\,,
$$

$$
\langle  \widetilde 0 \vert \widetilde\varphi (x^\pm) \widetilde\varphi (0) \vert \widetilde 0  \rangle = \,- \,\frac{i}{4}\, - \frac{1}{4 \pi} \ln [ i \mu( x^\pm + i x^0 \epsilon )]\,.
$$

\no One has the equal-time commutation relations 
$$
[\varphi (x) , \partial_0 \varphi (y)] = i\,\delta (x^1 - y^1)\,,
$$
$$
[\widetilde\varphi (x) , \partial_0 \widetilde\varphi (y)] =\,-\, i\,\delta (x^1 - y^1)\,.
$$

\no The canonical momenta are
$$
\pi (x) = \partial_0 \varphi (x)\,,
$$
$$
\widetilde\pi (x) = - \partial_0 \widetilde\varphi (x)\,,
$$

\no in such a way that the canonical equal-time commutation relations are given by
$$
[ \varphi (x), \pi (y)] = i \delta (x^1 - y^1)\,,
$$
$$
[ \widetilde\varphi (x), \widetilde\pi (y)] = i \delta (x^1 - y^1)\,.
$$

\no The dynamical equations are
$$
\partial_0 \varphi (x) = - \,i\,[ \varphi (x) , H ] = -\,i\,[\varphi (x) , \widehat H ]\,,
$$
$$
\partial_0 \widetilde\varphi (x) =  \,i\,[ \widetilde\varphi (x) , \widetilde H ] = -\,i\,[\widetilde\varphi (x) , \widehat H ]\,,
$$
where the total Hamiltonian $\widehat H = H - \widetilde H$ is the generator of time evolution of the combined system. Since the charges are defined as the space integral of the zero components of the currents, one has 

$$
[ {\cal Q}_e , \psi (x) ] = -\,\psi (x)\,\,,\,\,
[ {\cal Q}^5_e , \psi (x) ] = -\gamma^5\,\psi (x)\,,
$$ 
$$
[ \widetilde{\cal Q}_e , \widetilde\psi (x) ] = -\,\widetilde\psi (x)\,\,,\,\,
[ \widetilde{\cal Q}^5_e , \widetilde\psi (x) ] = -\gamma^5\,\widetilde\psi (x)\,,
$$ 

\no and analogously for the commutation relations of the phonon fields with ${\cal Q}_{ph}$ and $\widetilde{\cal Q}_{ph}$. This commutation relations are invariant under a unitary transformation and thus will be retained at finite temperature \cite{Ojima}. This explain the prescription for defining the ''tilde'' Wick-ordered exponential operators.

We shall work in the interaction picture in order to introduce temperature ($\beta = 1 / kT$) using the thermofield dynamics approach, and to compute the perturbative expansion for the phonon-phonon thermal correlation function. The thermal vacuum state $\vert 0 (\beta ) \rangle$ is given by \cite{TField,Ojima}

\be
\vert 0 (\beta ) \rangle = {\cal U} [ \theta ]\, \vert \widetilde 0 , 0 \rangle\,.
\ee

\no Denoting by $a^\dagger (p^1) \,(a (p^1 ))$ the creation (annihilation) operator for the bosonic field $\varphi (x)$ and by $b^\dagger (p^1) (b (p^1))$, the corresponding operators for the field $\eta (x)$, the unitary operator taking one to the bosonic thermofields is given by 

\be
{\cal U} [ \theta (\beta ) ]\,=\,e^{\,-\,\textstyle{\int_{- \infty}^{+ \infty } \Big ( [ \tilde a (p^1) a (p^1 ) \,-\,a^\dagger (p^1 ) \tilde a^\dagger (p^1) ]\,+\,[ \,\tilde b (p^1) b (p^1 ) \,-\,b^\dagger (p^1 ) \tilde b^\dagger (p^1)\, ] \Big ) \theta (\vert p^1 \vert ; \beta )\,dp^1}}\,,
\ee

\no and the Bogoliubov parameter $\theta ( \vert p^1 \vert , \beta )$ is implicitly defined by

\be
\sinh \theta (\vert p^1 \vert ; \beta ) = \frac{e^{\,-\,\beta \vert p^1 \vert / 2 }}{\sqrt{1\,-\,e^{\,-\,\beta\,\vert p^1 \vert}}}\,,
\ee

\be
\cosh \theta (\vert p^1 \vert ; \beta ) = \frac{1}{\sqrt{1\,-\,e^{\,-\,\beta\,\vert p^1 \vert}}}\,,
\ee  

\no with the Bose-Einstein statistical weight given by

\be
{\cal N} (\vert p^1 \vert ; \beta ) = \sinh^2 \theta (\vert p^1 \vert , \beta ) = \frac{1}{1\,-\,e^{\,-\,\beta\,\vert p^1 \vert}}\,.
\ee

\no The bosonic annihilation operators $d (p^1)$ and $\tilde d (p^1)$ transform according with (for more details see Refs. \cite{TField,ABR1,ABR2,ABR3})

\be
d ( p^1 ; \beta ) = {\cal U} [ - \theta (\beta ) ]\,d (p^1 )\,{\cal U} [ \theta (\beta ) ]\,=\,b (p^1) \cosh \theta (\vert p^1 \vert ; \beta ) - \tilde b^\dagger (p^1) \sinh \theta ( \vert p^1 \vert ; \beta )
\ee

\be
\tilde d ( p^1 ; \beta ) = {\cal U} [ - \theta (\beta ) ]\,\tilde d (p^1 )\,{\cal U} [ \theta (\beta ) ]\,=\,\tilde b (p^1) \cosh \theta (\vert p^1 \vert ; \beta ) -  b^\dagger (p^1) \sinh \theta ( \vert p^1 \vert ; \beta )\,.
\ee
\no The vacuum state at finite temperature satisfies
$$
d (p^1 ; \beta ) \vert 0 (\beta ) \rangle = 0\,\,\,,\,\,\,\tilde d (p^1 ; \beta ) \vert 0 (\beta ) \rangle = 0\,.
$$

\no Decomposing into creation and anihilation parts, at finite temperature one has

$$
\phi (x^\pm ; \beta ) = \phi^{(+)} (x^\pm ; \beta ) + \phi^{(-)} (x^\pm ; \beta )\,,
$$
$$
\widetilde\phi (x^\pm ; \beta ) = \widetilde\phi^{(+)} (x^\pm ; \beta ) + \widetilde\phi^{(-)} (x^\pm ; \beta )\,,
$$

\no where ($p = \vert p^1 \vert$)

\be
\pmatrix{\phi^{(+)} (x^\pm ; \beta ) \cr \widetilde\phi^{(+)} (x^\pm ; \beta )} = \int_0^\infty (dp) \pmatrix{ f_p (x^\pm) d (- p) \cosh \theta ( p ; \beta ) - f_p^\ast (x^\pm ) \tilde d (- p ) \sinh \theta ( p ; \beta ) \cr
f^\ast_p (x^\pm) \tilde d (- p) \cosh \theta ( p ; \beta ) - f_p (x^\pm ) d (- p ) \sinh \theta ( p ; \beta ) }\,,
\ee
\be
\pmatrix{\phi^{(-)} (x^\pm ; \beta ) \cr \widetilde\phi^{(-)} (x^\pm ; \beta )} = \int_0^\infty (dp) \pmatrix{ f^\ast_p (x^\pm) d^\dagger (+ p) \cosh \theta ( p ; \beta ) - f_p (x^\pm ) \tilde d^\dagger ( + p) \sinh \theta ( p ; \beta ) \cr
f_p (x^\pm) \tilde d^\dagger (+ p) \cosh \theta ( p ; \beta ) - f^\ast_p (x^\pm ) d^\dagger (+ p ) \sinh \theta ( p ; \beta ) }\,.
\ee

The thermofield Wick-ordered exponential $E (x ; \beta )$ of the free massless field $\phi (x ; \beta )$  is given by \cite{ABR1,ABR2,ABR3}

$$
E (x ; \beta ) = {\cal U} ( - \theta ) E (x) {\cal U} (\theta ) = e^{\textstyle\,- \frac{\lambda^2}{2 \pi} \it{z} (\mu , \beta )}\,\dt e^{\textstyle i \,\lambda\,\phi (x ; \beta )}\dt\,,
$$

\no (the same for the exponential of $\widetilde\phi$)  where $\it{z} (\mu ; \beta )$ is the infrared divergent integral \cite{ABR1,ABR3,ABR3} with $\mu$ an IR regulator. For a fixed finite temperature, the divergent terms in the asymptotical limit $\mu \rightarrow 0$ are

\be\label{z}
\it{z} (\mu ; \beta ) = \int_\mu^\infty \frac{dp}{p \big (e^{\beta\,p} - 1)}\,\approx \frac{\pi}{\beta\,\mu} + \frac{1}{2} \ln \big ( \frac{\beta \mu}{\pi} \big )  \equiv \it{f} (\mu , \beta ) + \frac{1}{2} \ln \big ( \frac{\beta \mu}{\pi} )\,.
\ee

Taking this into account, we can write the regularized thermal phonon fields (\ref{phi}) -(\ref{phitilde}) as 

\be\label{phit}
\pmatrix{\Phi (x ; \beta ) \cr \widetilde\Phi (x ; \beta ) } = \big ( \mu )^{\textstyle \frac{\alpha}{4 \pi}}\,e^{\textstyle\,-\,\frac{\alpha^2}{2 \pi}\,\it{z} (\mu , \beta )}\,\pmatrix{ \dt e^{\textstyle\,i\,\alpha\,\varphi (x ; \beta )} \dt \cr
\dt e^{\textstyle\,i\,\alpha\,\widetilde\varphi (x ; \beta )} \dt }\,=\, \Big (2 \beta \Big )^{\textstyle\,-\,\frac{\alpha^2}{4 \pi}}\,e^{\textstyle\,-\,\frac{\alpha^2}{2 \pi} \it{f} (\mu ; \beta)} \pmatrix{\dt e^{\textstyle\,i\,\alpha\, \varphi (x ; \beta)} \dt \cr
\dt e^{\textstyle\,i\,\alpha\, \widetilde\varphi (x ; \beta)} \dt}\,.
\ee

\no In the same way, the IR regularized thermal fermion chiral densities are

\be\label{J}
\pmatrix{J ( x ; \beta ) \cr \widetilde J (x ; \beta )} =  \big ( 2 \beta \big )^{\textstyle - \frac{\gamma^2}{4 \pi}} e^{\textstyle - \frac{\gamma^2}{2 \pi} \it{f} (\mu , \beta )}
\pmatrix{ \dt e^{\,\textstyle i \eta (x ; \beta )} \dt \cr
\dt e^{\,\textstyle i \widetilde\eta (x ; \beta )} \dt}\,.
\ee

At finite temperature the electron-phonon total interaction  Lagrangian is then given by 

\be\label{IL}
\widehat{\cal L}_I ( z ; \beta ) =   G^\prime  \Big [ \frac{\beta}{\pi} e^{\textstyle 2 \it{f}{ (\mu , \beta )}} \Big ]^{\textstyle -\,D} \Big \{ \dt \cos \big ( \gamma \eta (z ; \beta ) - \alpha \varphi (z ; \beta ) \big ) \dt -
\dt \cos \big ( \gamma \widetilde\eta (z ; \beta ) - \alpha \widetilde\varphi (z ; \beta ) \big ) \dt \Big \}\,,
\ee

\no with $G^\prime = G \big (4 \pi^2\big )^{ - \frac{\gamma^2}{4 \pi}}$ and

$$
D = \frac{(\alpha^2 + \gamma^2 )}{4 \pi}\,,
$$

\no is the scale dimension of the electron-phonon interaction term.

\section{ Thermal correlation function and the  electron-phonon selection rule}

In order to display the electron-phonon selection rule associated with the $\big ({\cal U}^5_e (1) \otimes {\cal U}_{ph}(1)\big ) \times \big (\widetilde{\cal U}^5_e (1) \otimes \widetilde{\cal U}_{ph}(1)\big )$ symmetry at finite temperature, we shall compute the thermal phonon two-point function. Since we are working in the interaction picture, the thermal 2-point function is given by the perturbative Gell'Mann-Low formula 
at finite temperature \cite{Ojima,ABR2,ABR3}
\be\label{2pf}
\langle 0 (\beta ) \vert T \Big ( \Phi (x) \Phi^\ast (y) \Big ) \vert 0 ( \beta ) \rangle = \frac{\textstyle
\langle 0 , \widetilde{0}  \vert T \Big ( \Phi (x ; \beta ) \Phi^\ast (y ; \beta )\,e^{\,\textstyle i\,\int\,\widehat{\cal L}_I (z ; \beta ) d^2 z} \Big ) \vert \widetilde{0} , 0 \rangle }{\textstyle \langle 0 , \widetilde{0}  \vert T e^{\,\textstyle i\,\int\,\widehat{\cal L}_I (z ; \beta ) d^2 z}  \vert \widetilde{0} , 0 \rangle}\,,
\ee

\no where $T (\cdots)$ means time-ordered product and the total interaction Lagrangian is given  by (\ref{IL}). We denote the thermal vacuum-vacuum amplitude by

\be\label{N}
N (\beta ; \mu ) \doteq \langle 0 , \widetilde{0}  \vert T e^{\,\textstyle i\,\int\,\widehat{\cal L}_I (z ; \beta ) d^2 z}  \vert \widetilde{0} , 0 \rangle\,.
\ee

\no The limit of the IR cutoff $\mu \rightarrow 0$ is performed at the end of all computations. As we shall see, the electron-phonon selection rule  enables one to obtain a finite perturbative expansion for the thermal two-point function independent of the IR cutoff 
\be
\lim_{\mu \rightarrow 0} \langle 0 (\beta ) \vert \Phi (x) \Phi^\ast (y ) \vert 0 (\beta ) \rangle \neq 0\,.
\ee

In order to have control on the origin of the cut-off dependent terms in the perturbative expansion we use the notation 

\be
\Phi ( x ; \beta ) =  \Big (2 \beta \Big )^{\textstyle\,-\,\frac{\alpha_x^2}{4 \pi}}\,e^{\textstyle\,-\,\frac{\alpha_x^2}{2 \pi} \it{f} (\mu ; \beta)}\,\Sigma (x , \alpha_x ;\beta )\,,
\ee  
\be
\widetilde\Phi (  y ; \beta ) =  \Big (2 \beta \Big )^{\textstyle\,-\,\frac{\alpha_y^2}{4 \pi}}\,e^{\textstyle\,-\,\frac{\alpha_y^2}{2 \pi} \it{f} (\mu ; \beta)}\,\widetilde\Sigma (y , \alpha_y ;\beta )\,,
\ee  
\be
\Sigma ( x , \alpha_x ; \beta ) = \dt e^{\textstyle i \alpha_x \varphi (x ; \beta )}\,,
\ee
\be
\widetilde\Sigma ( x , \alpha_x ; \beta ) = \dt e^{\textstyle i \alpha_x \widetilde\varphi (x ; \beta )}\,,
\ee
\be
W (z , \lambda ; \beta ) = \dt e^{\,\textstyle i\, \lambda\, \big ( \gamma\eta (z ; \beta ) - \alpha \varphi (z ; \beta \big )} \dt\,,
\ee
\be
\widetilde W (\tilde z , \tilde\lambda ; \beta ) = \dt e^{\,\textstyle  i\, \tilde\lambda\, \big ( \gamma \widetilde\eta (\tilde z ; \beta ) - \alpha \widetilde\varphi (\tilde z ; \beta )} \dt\,,
\ee

\no with  $\lambda, \tilde\lambda = \pm 1$. Using the fact that two thermal fields $\Phi (x ; \beta )$ and $\widetilde\Phi (y ; \beta )$ commute \footnote{The commutation relations at zero temperature are retained at finite temperature \cite{Ojima,ABR}.}, expanding the exponential of the interaction Lagrangian in  eq. (\ref{2pf}) in a power series of the electron-phonon coupling constant $G^\prime$ (se Appendix A for more details) we can write the general expression for the thermal phonon 2-point function as 

$$
\langle 0 (\beta ) \vert T \Phi (x) \Phi^\ast (y) \vert 0 (\beta ) \rangle = 
\frac{1}{N (\beta , \mu )} \sum_{n = 0}^\infty \big ( \frac{ 2 i G^\prime}{2} \big )^n \sum_{m, \tilde m} \frac{\delta_{m + \tilde m , n}}{m!\tilde m!} ( - 1 )^{\tilde m}\times
$$
\be\label{t2pf}
\int \prod_{\ell = 1}^m d^2 dz_\ell \int \prod_{k = 1}^{\tilde m} d^2 \tilde z_k\,\times
\sum_{\{\lambda_\ell\}_m} \sum_{\{\tilde{\lambda}_k\}_{\tilde m}} {\cal G}^{(m , \tilde m )} (z,\lambda , \tilde z , \tilde\lambda , D ; \beta )
{\cal F}^{(m , \tilde m )} (x , \alpha_x , y , \alpha_ y , z , \lambda , \tilde z , \tilde\lambda , \gamma ; \beta )\,,
\ee

\no where $\lambda_\ell, \tilde\lambda_k = \pm 1$, and $\sum_{\{\lambda_\ell\}_m } \big ( \sum_{\{\tilde \lambda_k\}_{\tilde m}}\big )$ runs over all possibilities in the set $\{\lambda_1, \cdots,\lambda_m\} \big ( \{ \tilde\lambda_1, \cdots ,\tilde\lambda_{k}\} \big )$ and
$$
{\cal G}^{(m , \tilde m )} (z,\lambda , \tilde z , \tilde\lambda , D ; \beta ) = \Big [ 2 \beta e^{\textstyle  \it{f} (\mu ; \beta )} \Big ]^{\, - D \textstyle 
\Big ( \sum_\ell^m \lambda_\ell - \sum_k^{\tilde m} \tilde \lambda_k \Big )^2} e^{\textstyle 2 i \pi D \sum_{k^\prime , k}^{\tilde m} \tilde\lambda_k \tilde\lambda_{k^\prime}}\times
$$
\be \label{GG}
\prod_{\ell^\prime > \ell}^m \Big [ \Omega (z_\ell - z_{\ell^\prime} ; \beta ) \Big ]^{\textstyle D \lambda_{\ell^\prime} \lambda_\ell}
\prod_{k^\prime > k}^{\tilde m} \Big [ \widetilde\Omega (z_k - z_{k^\prime} ; \beta ) \Big ]^{\textstyle D \tilde\lambda_{k^\prime} \tilde\lambda_k}
\prod_{\ell = 1}^m \prod_{k = 1}^{\tilde m} \Big [ \Omega (z_\ell - z_{k} - i \frac{\beta}{2} ) \Big ]^{\textstyle - D \lambda_\ell \tilde\lambda_k}\,,
\ee

$$
{\cal F}^{(m , \tilde m )} (x , \alpha_x , y , \alpha_ y , z , \lambda , \tilde z , \tilde\lambda , \gamma ; \beta ) = 
\Big ( 2 \beta \Big )^{\textstyle - \frac{(\alpha_x - \alpha_y )^2}{4 \pi}} e^{\textstyle -  \frac{(\alpha_x^2 - \alpha_y^2)}{2 \pi}\,\it{f} (\mu , \beta )} 
\Big [ \Omega (x - y ; \beta ) \Big ]^{\textstyle - \frac{\alpha_x \alpha_y}{4 \pi}} \times
$$
\be\label{FF}
\prod_{j = 1}^m \Big [ \Omega (x - z_j ; \beta ) \Big ]^{\textstyle - \frac{\alpha_x \alpha}{4 \pi} \lambda_j}
\prod_{k = 1}^{\tilde m} \Big [ \Omega (x - \tilde z_k - i \frac{\beta}{2} ) \Big ]^{\textstyle  \frac{\alpha_x \alpha}{4 \pi} \tilde\lambda_k}
\prod_{j = 1}^m \Big [ \Omega (y - z_j ; \beta ) \Big ]^{\textstyle - \frac{\alpha_y \alpha}{4 \pi} \lambda_j}
\prod_{k = 1}^{\tilde m} \Big [ \Omega (y - \tilde z_k - i \frac{\beta}{2} ) \Big ]^{\textstyle  \frac{\alpha_y \alpha}{4 \pi} \tilde\lambda_k}\,.
\ee

\no The terms in Eq. (\ref{GG}) are those that contribute to the thermal vacuum-vacuum amplitude $N (\beta , \mu )$ (\ref{N}) and in the limit $\mu \rightarrow 0$ the only non-zero contributions are those  with satisfy the selection rule

\be
\label{sr}
\sum_{\ell = 1}^m \lambda_\ell - \sum_{k = 1}^{\tilde m} \tilde\lambda_k = 0\,.
\ee

\no This result just manifests the conservation of the total electron-phonon charge generating the  $\big (U^5 (1)_e \otimes U (1)_{ph}\big ) \times \big (\widetilde U^5 (1)_e \otimes \widetilde U (1)_{ph}\big )$ composite symmetry of the interaction Lagrangian, which also implies for the thermal vacuum-vacuum amplitude 

$$
\lim_{\mu \rightarrow 0} N (\beta , \mu ) = N (\beta )\,.
$$

The $U_{ph}(1)$ invariant phonon two-point function corresponds to set $\alpha_x = \alpha_y = \alpha$, such that the Eq. (\ref{FF}) is finite, and the two-point function (\ref{t2pf}) is finite independent of the infrared cut-off $\mu$. However, if we set $\alpha_y = 0, \alpha_x = \alpha$, one formally obtains the thermal one-point function and in the limit $\mu \rightarrow 0$ one has

$$
\lim_{\mu \rightarrow 0} \langle 0 (\beta ) \vert \Phi (x) \vert 0 (\beta ) \rangle = 0\,,
$$

\no implying that the $U_{ph} (1)$ symmetry is not broken, which is in accordance with the cluster decomposition property

$$
\lim_{{\zeta \rightarrow \infty}\atop{\zeta^2 < 0}} \langle 0 (\beta ) \vert \Phi (x) \Phi^\ast (x + \zeta ) \vert 0 (\beta ) \rangle = \big\vert\big\vert \langle 0 (\beta ) \vert \Phi (x) \vert 0 (\beta ) \rangle  \big\vert\big\vert^2 = 0\,.
$$ 

In the same way one may compute the general mixed correlation functions involving the phonon fields ($\Phi$,$\widetilde\Phi$)  and the electron chiral densities ($J$,$\widetilde J$).

\section{Concluding Remarks}
\setcounter{equation}{0}

We used the Thermofield Dynamics and the Thermofield bosonization to discuss the symmetry aspects of the Lagrangian model of the ICDW by assuming a dynamical phase phonon approach. This leads to thermal periodic sine-Gordon potential at finite temperature. This potential is responsible for the mass gap generation, a phenomenon well known for long time to occur in several two-dimensional quantum field models, such as the chiral Gross-Neveu model \cite{AAR,G-N,Col}. 

The dynamical generation of a mass gap in the ICDW can be viewed as follows: Rename the fields 
$$
\eta = \phi_1\,,\,\widetilde\eta = \widetilde\phi_1\,,
$$
$$\varphi = \phi_2, \widetilde\varphi_2 = \widetilde\phi_2\,,
$$

\no  introducing a new Fermi field specie $\psi^2,\widetilde\psi^2$ in terms of the Mandelstam representation (\ref{psi}) for $\phi_2$ and $\widetilde\phi_2$, the fermion operators (\ref{psi}) are $\psi = \psi^1$ and $\widetilde\psi = \widetilde\psi^1$ and  the phonon fields
can be written in terms of a chiral condensate of the fermions $\psi^2$ and $\widetilde\psi^2$  
$$
\Phi (x) =  \dt \psi^{2 \dagger}_{\ell} (x)\psi^2_r (x) \dt\,,
$$
$$
\widetilde\Phi (x) =  \dt \widetilde\psi^{2 \dagger}_{\ell} (x)\widetilde\psi^2_r (x) \dt\,.
$$

\no The conserved currents generating the extended linked electron-phonon symmetry can be written as ($\alpha = \gamma = 2 \sqrt\pi$)
$$
{\cal I}^\mu (x) = j^{\mu 5}_e (x) + j^\mu_{ph} (x) = - \frac{1}{\sqrt \pi} \partial^\mu \big ( \phi^1 (x) + \phi^2 (x) \big ) = - \sqrt{\frac{2}{\pi}} \partial^\mu \phi^+ (x)\,,
$$
$$
\widetilde{\cal I}^\mu (x) = \tilde j^{\mu 5}_e (x) + \tilde j^\mu_{ph} (x) =  \frac{1}{\sqrt \pi} \partial^\mu \big ( \widetilde\phi^1 (x) + \widetilde\phi^2 (x) \big ) = \sqrt{\frac{2}{\pi}} \partial^\mu \widetilde\phi^+ (x)\,,
$$

\no with $\partial_\mu {\cal I}^\mu (x) = \partial^2 \phi^+ (x) = 0$, the same for the tilde current, where we have introduced the independent fields
$$
\phi^\pm (x) = \frac{1}{\sqrt 2} \big ( \phi^1 (x) \pm \phi^2 (x) \big )\,,
$$
$$
\widetilde\phi^\pm (x) = \frac{1}{\sqrt 2} \big ( \widetilde\phi^1 (x) \pm \widetilde\phi^2 (x) \big )\,.
$$

\no In terms of these fields, the bosonized effective Lagrangians (\ref{LBos})-(\ref{tLBos}) at finite temperature can be written as
$$
{\cal L} (x ; \beta ) = \frac{1}{2} \dt \big ( \partial_\mu \phi^+ (x ; \beta ) \big )^2 \dt +  \frac{1}{2} \dt \big ( \partial_\mu \phi^- (x ; \beta ) \big )^2 \dt + G \big ( \frac{\mu}{2 \pi } \big ) \dt \cos \big ( \sqrt{8 \pi} \phi^- (x ; \beta ) \big )\dt\,..
$$
$$
\widetilde{\cal L} (x ; \beta ) = \frac{1}{2} \dt \big ( \partial_\mu \widetilde\phi^+ (x ; \beta ) \big )^2 \dt +  \frac{1}{2} \dt \big ( \partial_\mu \widetilde\phi^- (x ; \beta ) \big )^2 \dt + G \big ( \frac{\mu}{2 \pi } \big ) \dt \cos \big ( \sqrt{8 \pi} \widetilde\phi^- (x ; \beta ) \big )\dt\,,
$$
\no which corresponds to the bosonized version of the chiral Gross-Neveu model with two fermion species \cite{G-N,BAQ1} generalized for finite temperature
$$
{\cal L}_{GN} (x ; \beta )  = \sum_{j = 1}^2 \dt\bar\psi^j (x ; \beta )  i \gamma^\mu \partial_\mu \psi^j (x ; \beta )\dt\, + 
$$
$$
\frac{g^2}{2} \big \{\dt (\bar\psi^1 (x ; \beta )\psi^1 (x ; \beta) )(\bar\psi^2(x ; \beta )\psi^2 (x  \beta ))\dt - \dt(\bar\psi^1 (x ; \beta )\gamma^5\psi^1 (x ; \beta ))(\bar\psi^2 (x ; \beta )\gamma^5\psi^2 (x ; \beta ))\dt \big \}\,,
$$
\no and similary for $\widetilde{\cal L}_{GN} (x ; \beta )$. In terms of spinor components the GN interaction Lagrangian can be written as

\be\label{GNIL}
{\cal L}_{GN}^I (x ; \beta ) = g^2 \big \{ \dt\psi_r^{1 \ast} (x ; \beta )\psi_\ell^1 (x ; \beta) \psi_\ell^{2 \ast} (x ; \beta )\psi_r^2 (x  \beta )\dt + 
\dt\psi_r^{2 \ast} (x ; \beta )\psi_\ell^2 (x ; \beta) \psi_\ell^{1 \ast} (x ; \beta )\psi_r^1 (x  \beta )\dt \big \}\,.
\ee

In terms of the independent Bose fields $\phi^\pm$ ($\widetilde\phi^\pm$) the Fermi fields $\psi^j$ ($\widetilde\psi^j$) with Lorentz spin $1 / 2$ can be decomposed in terms of two spin $1 / 4$ components
$$
\psi^j (x ; \beta ) = S (x ; \beta )\,\hat\psi^j (x ; \beta )\,,
$$ 
$$
\widetilde\psi ^j (x ; \beta ) = \widetilde S (x ; \beta )\,\hat{\widetilde\psi^j} (x ; \beta )\,,
$$ 

\no where the linked electron-phonon selection rule is carried by the fields 

$$
S (x ; \beta ) = \Big ( \frac{\mu}{2 \pi } \Big )^{\frac{1}{4}}\,\dt e^{\,\textstyle i \sqrt{\frac{\pi}{2}} \{ \gamma^5 \phi^+ (x ; \beta ) + \int_{x^1}^\infty \partial_0 \phi^+ (x^0 , z^1 ; \beta ) d z^1\} }\,,
$$
$$
\widetilde S (x ; \beta ) = \Big ( \frac{\mu}{2 \pi } \Big )^{\frac{1}{4}}\,\dt e^{\,\textstyle i \sqrt{\frac{\pi}{2}} \{ \gamma^5 \widetilde\phi^+ (x ; \beta ) + \int_{x^1}^\infty \partial_0 \widetilde\phi^+ (x^0 , z^1 ; \beta ) d z^1\} }\,,
$$

\no and the components $\hat\psi^j$, $\hat{\widetilde\psi} ^j$ are given by ($ j = 1,2.$)
$$
\hat\psi ^j (x ; \beta ) = \Big ( \frac{\mu}{2 \pi } \Big )^{\frac{1}{4}}\,\dt e^{\, \textstyle  \, (- 1 )^{j + 1}\,i\, \sqrt{\frac{\pi}{2}} \{ \gamma^5 \phi^- (x ; \beta ) + \int_{x^1}^\infty \partial_0 \phi^- (x^0 , z^1 ; \beta ) d z^1 \}}\,,
$$
$$
\hat{\widetilde\psi^j} (x ; \beta ) = \Big ( \frac{\mu}{2 \pi } \Big )^{\frac{1}{4}}\,\dt e^{\, \textstyle \, (- 1 )^{j + 1}\,i\, \sqrt{\frac{\pi}{2}} \{ \gamma^5 \widetilde\phi^- (x ; \beta ) + \int_{x^1}^\infty \partial_0 \widetilde\phi^- (x^0 , z^1 ; \beta ) d z^1 \}}\,,
$$

\no with the property
$$
\hat\psi^1 = \hat\psi^{2 \ast} \equiv \hat\psi\,.
$$
$$
\hat{\widetilde\psi^1}  = \hat{\widetilde\psi}^{2 \ast} \equiv \hat{\widetilde\psi}\,.
$$

\no This enables the definition of the fields $\Psi$ and $\widetilde\Psi$ with spinor components
$$
\Psi =  \pmatrix{\Psi_\ell \cr \Psi_r } \equiv \pmatrix{\dt\hat\psi_\ell\hat\psi_\ell\dt\cr \dt \hat\psi_r\psi_r \dt}\,,
$$

\no such that the electron-phonon condensate corresponding to the interaction piece  can be written as a mass term for the field $\Psi$

$$
{\cal L}_I (x ; \beta ) = G \Big ( \dt \Psi^\ast_r (x ; \beta )\Psi_\ell (x ; \beta ) \dt + \dt \Psi_\ell^\ast (x ; \beta )\Psi_r (x ; \beta )\dt \Big )  = G \bar\Psi (x ; \beta )\Psi (x ; \beta)\,,
$$
\no and the same for $\widetilde{\cal L}_I (x ; \beta )$. Within the bosonized version of the models and under the dynamical phonon-phase approximation one has the mapping
\be
{\cal L}_{e-ph} (x ; \beta ) \equiv {\cal L}_{GN} (x ; \beta )\,.
\ee
\no The electron-phonon symmetry corresponds to the $U (1) \otimes U^5 (1)$ symmetry of the underlying Gross-Neveu model and one has

$$
\langle 0 (\beta ) \vert \bar\Psi (x)\Psi (x)  \vert 0 (\beta ) \rangle \neq 0\,.
$$

\no The fact that within the dynamical phonon-phase approach the phonon field ''behaves'' like a chiral electron density enables the appearance of the dynamical energy gap generation even for finite temperature.

In order to conclude, let us make a remark on the introduction of an external electric field
$$
E (x) = \frac{1}{2} \epsilon^{\mu \nu}{\cal F}_{\mu \nu} (x) = \epsilon^{\mu \nu} \partial_\mu A_\nu (x)\,.
$$

\no The external field couples with the electronic field as
$$
{\cal L}^{\prime\prime}(x ; \beta ) = -e A_\mu (x)\dt\bar\psi(x;\beta) \gamma^\mu \psi (x;\beta )  = \frac{e}{\sqrt\pi} \eta (x ; \beta ) \epsilon^{\mu\nu}\partial_\nu A_\mu (x)\,,
$$  
$$
\widetilde{\cal L}^{\prime\prime}(x ; \beta ) = -e A_\mu (x) \bar{\widetilde\psi} (x ; \beta )\gamma^\mu \widetilde\psi (x;\beta) = - \frac{e}{\sqrt\pi} \widetilde\eta (x ; \beta ) \epsilon^{\mu\nu}\partial_\nu A_\mu (x)\,.
$$

\no In terms of the fields $\phi^\pm$ one has
$$
{\cal L}^{\prime\prime}(x ; \beta ) = {e}\sqrt{\frac{2}{\pi}} \{\phi^+ (x ; \beta ) + \phi^- (x;\beta ) \} \epsilon^{\mu\nu}\partial_\nu A_\mu (x)\,.
$$

\no The field $\phi^+$ is no longer a free field and one has the current acceleration equation 
$$
\partial_\mu {\cal I}^\mu (x ;\beta ) = - \frac{1}{\sqrt\pi} \partial^2 \phi^+ (x;\beta ) = \frac{e}{\pi} E (x)\,.
$$

\no At finite temperature, although the total Lagrangian $\widehat{\cal L}^{\prime\prime}( ; \beta )$ also exhibits the extended linked symmetry  under the linked $(U^5 (1)_e\otimes U (1)_{ph} ) \times (\widetilde U^5(1)_e \otimes \widetilde U (1)_{ph} )$ transformations, the presence of the external electric potential shares an anomalous effect analogously to that of zero temperature case. The dynamical mass gap generation and the anomaly effect remain at finite temperature.

{\bf Acknowledgements}: The authors are  grateful to Brazilian Research Council (CNPq) for partial financial support. The work of R. L. P. G. Amaral was supported by Coordenacao de Aperfeicoamento do Pessoal do Ensino Superior, bolsista CAPES, Proc.  BEX 0885/11-8.

\appendix{{\centerline{\bf{Appendix A}}}}
\centerline{\bf {Perturbative Expansion}}
\renewcommand{\theequation}{{A}.\arabic{equation}}\setcounter{equation}{0}

Expanding the exponential of the interaction Lagrangian in  eq. (\ref{2pf}) in a power series of the electron-phonon coupling constant $G^\prime$, one has

$$
e^{\textstyle\,i\,\int d^2 z \widehat{\cal L}_I (z ; \beta )} = 
$$
$$
\exp \Bigg ( \,i\,
G^\prime  \Big [ \frac{\beta}{\pi} e^{\textstyle 2 \it{f}{ (\mu , \beta )}} \Big ]^{\textstyle -\,D} \Big \{ \int d^2 z \dt \cos \big ( \gamma \eta (z ; \beta ) - \alpha \varphi (z ; \beta ) \big ) \dt -
\int d^2 \tilde z \dt \cos \big ( \gamma \widetilde\eta (z ; \beta ) - \alpha \widetilde\varphi (z ; \beta ) \big ) \dt \Big \} \Bigg ) =
$$
$$
\sum_{n = 0}^\infty \frac{(i G^\prime )^n}{n !}
 \Big [ \frac{\beta}{\pi} e^{\textstyle 2 \it{f}{ (\mu , \beta )}} \Big ]^{\textstyle -\,n D} \Big \{ \int d^2 z \dt \cos \big ( \gamma \eta (z ; \beta ) - \alpha \varphi (z ; \beta ) \big ) \dt -
\int d^2 \tilde z \dt \cos \big ( \gamma \widetilde\eta (z ; \beta ) - \alpha \widetilde\varphi (z ; \beta ) \big ) \dt \Big \}^n =
$$
$$
\sum_{n = 0}^\infty \frac{( i G^\prime )^n}{n !} \Big [ \frac{\beta}{\pi} e^{\textstyle 2 \it{f}{ (\mu , \beta )}} \Big ]^{\textstyle -\,n D}
\sum_{m, \tilde m} \frac{(n !) \delta_{m + \tilde m , n}}{m ! \tilde m !} ( - 1 )^{\tilde m} \Big ( \int d^2 z \dt \cos \big ( \gamma \eta (z ; \beta ) - \alpha \varphi (z ; \beta ) \big ) \dt \Big )^m \times
$$
$$
\Big (\int d^2 \tilde z \dt \cos \big ( \gamma \widetilde\eta (\tilde z ; \beta ) - \alpha \widetilde\varphi (\tilde z ; \beta ) \big ) \dt \Big )^{\tilde m} = \sum_{n = 0}^\infty ( i G^\prime )^n \Big [ \frac{\beta}{\pi} e^{\textstyle 2 \it{f}{ (\mu , \beta )}} \Big ]^{\textstyle -\,n D}
\sum_{m, \tilde m} \frac{\delta_{m + \tilde m , n}}{m ! \tilde m !} ( - 1 )^{\tilde m} \times
$$
$$
\int \prod_{\ell = 1}^m d^2 z_\ell\,
\int \prod_{k = 1}^{\tilde m} d^2 \tilde z_k\,\prod_{\ell = 1}^m \dt \cos \big ( \gamma \eta (z_\ell ; \beta ) - \alpha \varphi (z_\ell ; \beta ) \big ) \dt \prod_{k = 1}^{\tilde m}\dt \cos \big ( \gamma \widetilde\eta (\tilde z_k ; \beta ) - \alpha \widetilde\varphi (\tilde z_k ; \beta ) \big ) \dt =
$$
$$
\sum_{n = 0}^\infty \Big ( \frac{i G^\prime}{2} \Big )^n \Big [ \frac{\beta}{\pi} e^{\textstyle 2 \it{f}{ (\mu , \beta )}} \Big ]^{\textstyle -\,n D}
\sum_{m, \tilde m} \frac{\delta_{m + \tilde m , n}}{m ! \tilde m !} ( - 1 )^{\tilde m}
\int \prod_{\ell = 1}^m d^2 z_\ell\,
\int \prod_{k = 1}^{\tilde m} d^2 \tilde z_k \times
$$
\be
\sum_{\{\lambda_\ell\}_m }\sum_{\{\tilde \lambda_k\}_{\tilde m}} \prod_{\ell = 1}^m W (z_\ell ; \beta, \lambda_\ell) \prod_{k = 1}^{\tilde m} \widetilde{W} (\tilde z_k ; \beta, \tilde\lambda_k )\,,
\ee

\no where $\lambda_\ell, \tilde\lambda_k = \pm 1$, and $\sum_{\{\lambda_\ell\}_m } \big ( \sum_{\{\tilde \lambda_k\}_{\tilde m}}\big )$ runs over all possibilities in the set $\{\lambda_1, \cdots,\lambda_m\} \big ( \{ \tilde\lambda_1, \cdots ,\tilde\lambda_{k}\} \big )$.

\no The 2-point function is now

$$
\langle 0 (\beta ) \vert T \Big ( \Phi (x ; \beta ) \Phi^\ast (y ; \beta ) \Big ) \vert \ 0 (\beta ) \rangle \,=
$$
$$
\frac{1}{N (\beta , \mu )}\,\Big ( 2 \beta  \Big )^{\,\textstyle - \frac{(\alpha_x^2 + \alpha_y^2 )}{4 \pi}}e^{\textstyle \,-\,\frac{(\alpha_x^2 + \alpha_y^2)}{2\pi}\,\it{f} ( \mu , \beta )} \,\sum_{n = 0}^{\infty}\,\big ( i 2 G^\prime\big )^n \big ( 2 \pi \big )^{\,\textstyle -\,\frac{\gamma^2}{4 \pi} n} \times
\Big [ \frac{\beta}{\pi} e^{\,\textstyle 2 \it{f} (\mu , \beta )} \Big ]^{\,\textstyle\,-  n D}\,\frac{\delta_{m + \tilde m , n}}{m ! \tilde m !} ( - 1 )^{\widetilde m} \times
$$
\be \int \prod_{\ell = 1}^m d^2 z_\ell \int \prod_{k = 1}^{\widetilde m} d^2 \tilde z_k \sum_{\{\lambda_\ell\}_m }\sum_{\{\tilde \lambda_k\}_{\tilde m}} 
\langle 0 , \widetilde 0 \vert T \Big (\Sigma (x ; \beta ) \Sigma^\ast (y ; \beta ) \prod_{j = 1}^m\, W (z_j, \beta , \lambda_j ) \prod_{k = 1}^{\tilde m}\widetilde W ( \tilde z_k , \beta, \tilde\lambda_k )\Big ) \vert \widetilde 0 , 0 \rangle\,.
\ee

\no The Wick's theorem can be extended to generalized time-ordered product of Wick-ordered exponentials and we obtain
($\langle \cdots \rangle  \equiv \langle 0 , \widetilde 0 \vert  \cdots \vert \widetilde 0 , 0 \rangle$)
$$
\langle 0 , \widetilde 0 \vert T \Big (\Sigma (x ; \beta ) \Sigma^\ast (y ; \beta ) \prod_{j = 1}^m\, W (z_j, \beta , \lambda_j ) \prod_{k = 1}^{\tilde m}\widetilde{W} ( \tilde z_k , \beta, \tilde\lambda_k )\Big ) \vert \widetilde 0 , 0 \rangle =
$$
$$
\langle T \big ( \Sigma (x ; \beta ) \Sigma^\ast (y ; \beta ) \big ) \rangle \prod_{j = 1}^m \langle T \big ( \Sigma (x ; \beta ) W (z_j ; \beta , \lambda_j ) \big ) \rangle \prod_{k = 1}^{\tilde m} \langle T \big ( \Sigma (x ; \beta ) \widetilde{W} (\tilde z_k ; \beta , \tilde \lambda_k ) \big ) \rangle \times
$$
$$
\prod_{j = 1}^m \langle T \big ( \Sigma^\ast (y ; \beta ) W (z_j ; \beta , \lambda_j ) \big ) \rangle \prod_{k = 1}^{\tilde m} \langle T \big ( \Sigma^\ast (y ; \beta ) \widetilde{W} (\tilde z_k ; \beta , \tilde \lambda_k ) \big ) \rangle \times
$$
$$
\prod_{\ell^\prime > \ell}^m \langle T \big ( W (z_\ell ; \beta , \lambda_\ell ) W (z_{\ell^\prime} ; \beta , \lambda_{\ell^\prime} ) \big ) \rangle
\prod_{k^\prime > k}^{\tilde m}\langle T \big ( \widetilde{W} (\tilde z_k ; \beta , \tilde\lambda_k ) \widetilde{W} (\tilde z_{k^\prime} ; \beta , \tilde\lambda_{k^\prime} ) \big ) \rangle \times
$$
\be
\prod_{j = 1}^m \prod_{k = 1}^{\tilde m} \langle T \big ( W (z_j ; \beta , \lambda_j ) \widetilde{W} (\tilde z_k ; \beta , \tilde\lambda_k ) \big ) \rangle\,.
\ee

We use that

$$
\langle T \big ( \dt e^{\textstyle\,i a \Phi (x ; \beta)} \dt\,\dt e^{\,\textstyle i b \Phi (y ; \beta)} \dt \big ) \rangle = e^{ \textstyle - a b \langle T \Phi (x ; \beta ) \Phi (y ; \beta ) \rangle}\,,
$$

\no and the propagators of the scalar thermofields can be written \cite{ABR1,ABR2} as follows (use was made of (\ref{z}))

$$
\langle 0 , \widetilde 0 \vert T \Phi (x ; \beta ) \Phi (y ; \beta ) \vert \widetilde 0 , 0 \rangle = \frac{1}{\pi} \it{z} (\mu , \beta ) - \frac{1}{2 \pi} \ln \Big (\frac{\mu}{2 \pi}\Big ) - \frac{1}{4 \pi} \ln \Omega ( x - y ; \beta ) 
$$
\be
= \frac{1}{2\pi} \it{f} (\mu , \beta ) + \frac{1}{2 \pi} \ln (2 \beta ) - \frac{1}{4 \pi} \ln \Omega ( x - y ; \beta )\,,
\ee
$$
\langle 0 , \widetilde 0 \vert T \widetilde\Phi (\tilde x ; \beta ) \widetilde\Phi (\tilde y ; \beta ) \vert \widetilde 0 , 0 \rangle = - \frac{i}{2} + \frac{1}{\pi}\it{z} (\mu , \beta ) - \frac{1}{2 \pi} \ln \Big (\frac{\mu}{2 \pi}\Big ) - \frac{1}{4 \pi} \ln \widetilde\Omega ( \tilde x - \tilde y ; \beta ) 
$$
\be
= - \frac{i}{2} + \frac{1}{2\pi} \it{f} (\mu , \beta ) + \frac{1}{2 \pi} \ln (2 \beta ) - \frac{1}{4 \pi} \ln \widetilde\Omega ( \tilde x - \tilde y ; \beta )\,,
\ee
\be
\langle 0 , \widetilde 0 \vert T \Phi (x ; \beta ) \widetilde\Phi (\tilde y ; \beta ) \vert \widetilde 0 , 0 \rangle =
 - \frac{1}{2\pi} \it{f} (\mu , \beta ) - \frac{1}{2 \pi} \ln (2 \beta ) + \frac{1}{4 \pi} \ln \Omega (  x - \tilde y - i \frac{\beta}{2} )\,,
\ee

\no where the space-time dependent functions are given by

\be
\Omega (x - y ; \beta ) = (2 i \beta )^2 \sinh \Big [\frac{\pi}{\beta} \Big ( x^+ - y^+ - i \varepsilon (x^0 - y^0) \Big ) \Big ]
\sinh \Big [\frac{\pi}{\beta} \Big ( x^- - y^- - i \varepsilon (x^0 - y^0) \Big ) \Big ]
\ee

\be
\widetilde\Omega (\tilde x - \tilde y ; \beta ) = (2 i \beta )^2 \sinh \Big [\frac{\pi}{\beta} \Big ( \tilde x^+ - \tilde y^+ + i \varepsilon (\tilde x^0 - \tilde y^0) \Big ) \Big ]
\sinh \Big [\frac{\pi}{\beta} \Big ( \tilde x^- - \tilde y^- + i \varepsilon (x^0 - y^0) \Big ) \Big ]
\ee

\be
\Omega (x - \tilde y - i \frac{\beta}{2} ) = (2 i \beta )^2 \sinh \Big [\frac{\pi}{\beta} \Big ( x^+ - \tilde y^+ - i \frac{\beta}{2} \Big ) \Big ]
\sinh \Big [\frac{\pi}{\beta} \Big ( x^- - \tilde y^- - i \frac{\beta}{2} \Big ) \Big ]\,.
\ee

\no Using that 
$$
\sum_{\ell^\prime > \ell}^m \lambda_\ell \lambda_{\ell^\prime} + \sum_{k^\prime > k}^{\tilde m} \tilde\lambda_k \tilde\lambda_{k^\prime} - \sum_\ell^m \lambda_\ell \sum_k^{\tilde m} \tilde\lambda_k = \frac{1}{2} \Big ( \sum_\ell^m \lambda_\ell - \sum_k^{\tilde m} \tilde \lambda_k \Big )^2 - \frac{n}{2}\,,
$$

\no we obtain the expression (\ref{t2pf}) for the phonon two-point function.



\begin{thebibliography}{20}
\bibitem{AAR} E. Abdalla, M. C. Abdalla and K. D. Rothe, 'Non-perturbative Methods in 2 Dimensional Quantum Field Theory' (World Scientific, Singapore, 1991; Ibdem, 2nd Revised Version, 2000), and references quoted therein.
\bibitem{Fradkin} E. Fradkin, {Field Theories of Condensed Matter Physics}, Frontiers in Physics (Addison-Wesley, reading, MA, 1991, and references quoted therein.
\bibitem{Jackiw} R. Jakiw and J. R. Schrieffer, Nucl. Phys. {\bf B190}, 263 (1981).
\bibitem{Gruner} G . Gr$\ddot{u}$ner, Rev. Mod. Phys. {\bf 60}, No 4, (1988) 1129.
\bibitem{SakitaShizuya} B. Sakita and K. Shizuya, Phys. Rev. {\bf 42} (1990) 4486.
\bibitem{Saka2} B. Sakita and K. Shizuya, Phys. Lett. {\bf A 145}, 209 (1990);\\
B. Sakita and K. Shizuya, Phys. Rev. {\bf B 42} (1990) 5586;
\bibitem{Rice} P. M. Lee, T. M. Rice and T. W. Anderson, Solid State Comm. {\bf 14} (1974) 703.
\bibitem{Krive} I. V. Krive, A. S. Rozhavsky, Phys. Lett. {\bf A 113}, 313 (1985).
\bibitem{SuSakita}  B. Sakita and Z. -B. Su, Prog. Theor.Phys. Suppl {\bf 86}, 238 (1986); Z. B. Su and B. Sakita, Phys. Rev. Lett. {\bf 56} 780 (1986); Phys. Rev {\bf B 38}, 7241 (1988).
\bibitem{Ishikawa} M. Ishikawa and H. Takayama, Prog. Theo. Phys. {\bf 79}, 359 (1983).
\bibitem{Krive2} I. V. Krive, A. S. Rozhavsky, E. R. Mucciolo and L. E. Oxman, Phys. Rev. {\bf B 61}, 12835 (2000).
\bibitem{BAQ1} L. V. Belvedere, R. L. P. G. Amaral and A. F. Queiroz, Int. Journ. Modern Physics B, 16, No 31 (2002) 4685.
\bibitem{BAQ2} L. V. Belvedere, R. L. P. G. Amaral and A. F. Queiroz, Int. Jour. Modern Phys. B, 18, No 6 (2004) 883.
\bibitem{BAQ3} L. V. Belvedere, R. L. P. G. Amaral and A. F. Queiroz, Phys. Lett. {\bf A 289}, 177 (2001).
\bibitem{Sasaki} K. Sasaki , Phys. Rev.  {\bf B 65} (2002) 155429;
\bibitem{Hayashi} M.Hayashi and H. Youshioka, Phys. Rev. Lett 77 (1996) 3403..
\bibitem{Ya} Victor M. Yakovenko and Hsi-Sheng Goan, Phys. Lett. {\bf B 58} (1998) 10648.
\bibitem{Aperis} Aperis et al., Phys. Lett. {\bf B 702} (2011) 181.
\bibitem{Hind} M Hindmarsh, Physica {\bf B 178} (1992) 47.
\bibitem{Froh} H. Fr$\ddot{o}$hlich , Proc. R. Soc. London Ser. {\bf A 223} (1954) 296.
\bibitem{TField} H. Umezawa, H. Matsumoto and M. Tachiki, 1982, Thermofield Dynamics and Condensed States (Amsterdam, North-Holland).
\bibitem{TField1} L. Leplae, K. Mancini, H. Umezawa,Phys. Rep. 10 C (1974) 151.
\bibitem{TField2} Y. Takahashi, H. Umezawa, Collet. Phenomena 2 (1975) 55.
\bibitem{TField3} A. Das, Finite Temperature Field Theory, World Scientific, 1997.
\bibitem{Ojima} I. Ojima, Ann. Phys. 137 (1981) 1.
\bibitem{ABR1} R. L. P. G. Amaral, L. V. Belvedere and K. D. Rothe, Annals of Phys. {\bf 320} (2005) 299;
\bibitem{ABR2} R. L. P. G. Amaral, L. V. Belvedere and K. D. Rothe, Annals of Phys. {\bf 323} (2008) 2662;
\bibitem{ABR3} L. V. Belvedere, R. L. P. G. Amaral and K. D. Rothe, Journ. Phys. A, Math. and Theoretical {\bf 44} (2009) 015401;
\bibitem{Swieca} J. A. Swieca, Fortschr. Physik {\bf 25}, 303 (1977)
\bibitem{Col} S. Coleman, Phys. Rev. {\bf D 11}, 2088 (1975)
\bibitem{M} S. Mandelstam, Phys. Rev. {\bf D 11}, 3026 (1975)
\bibitem{G-N} D. J. Gross and A. Neveu, Phys. Rev. {\bf D10} (1974) 3235.
\end{thebibliography}
\end{document}